\documentstyle[psfig]{l-aa}
\begin{document}

\def\Lya{Ly$\alpha$}
\def\Lyal{Ly$\alpha$ $\lambda$1216}
\def\Hb{H$\beta$}
\def\nv{N\,{\sc v}}
\def\nvl{N\,{\sc v} $\lambda$1240}
\def\civ{C\,{\sc iv}}
\def\civl{C\,{\sc iv} $\lambda$1549}
\def\heiil{He\,{\sc ii} $\lambda$1640}
\def\oiiil{O\,{\sc iii}\,] $\lambda$1663}
\def\ciii{C\,{\sc iii}\,]}
\def\ciiil{C\,{\sc iii}\,] $\lambda$1909}
\def\Lrest{$\lambda\dmrm{rest}$}
\def\Lobs{$\lambda\dmrm{obs}$}
\def\kms{\mrm{km\,s^{-1}}}
\def\Hubble{\mrm{km\,s^{-1}\,Mpc^{-1}}}
\def\ergsA{\mrm{erg\,s^{-1}\,\mbox{\AA}^{-1}}}
\def\ergscm{\mrm{erg\,s^{-1}\,cm^{-2}}}
\def\ergscmA{\mrm{erg\,s^{-1}\,cm^{-2}\,\mbox{\AA}^{-1}}}
\def\Sig{\mbox{\large $\sigma$}}
\def\Et{\mbox{\large $\eta$}}
\def\tot{\dmrm{tot}}
\def\prin{\dmrm{prin}}
\def\rest{\dmrm{rest}}
\def\cont{\dmrm{cont}}
\def\lcont{\dmrm{\lambda,cont}}
\def\line{\dmrm{line}}
\def\lt{\dmrm{lt}}
\def\var{\dmrm{var}}
\def\ltsim{\raisebox{-.5ex}{$\;\stackrel{<}{\sim}\;$}}
\def\gtsim{\raisebox{-.5ex}{$\;\stackrel{>}{\sim}\;$}}
\newcommand{\mrm}[1]{\ifmmode \mathrm{#1} \else $\mathrm{#1}$\fi}
\newcommand{\dmrm}[1]{_{\mathrm{\,#1}}}
\newcommand{\umrm}[1]{^{\mathrm{\,#1}}}
\newcommand{\mean}[1]{\ifmmode \langle\,#1\,\rangle \else $\langle\,#1\,\rangle$\fi}

\thesaurus{02.12.3; 11.01.2; 11.17.2; 11.19.1; 13.21.1}

\title{Principal component analysis of two ultraviolet emission-lines in 18 active galactic nuclei}
\author{M. Türler \inst{1,2} \and T. J.-L. Courvoisier \inst{1,2}}
\institute{Geneva Observatory, ch. des Maillettes 51, CH-1290 Sauverny, Switzerland \and \textit{INTEGRAL} Science Data Centre, ch. d'Écogia 16, CH-1290 Versoix, Switzerland}
\offprints{M. Türler}
\date{Received XXX / Accepted YYY}
\maketitle
\markboth{M. Türler \& T. J.-L. Courvoisier: PCA of two UV emission-lines in 18 AGN}{M. Türler \& T. J.-L. Courvoisier: PCA of two UV emission-lines in 18 AGN}

\begin{abstract}
We apply a principal component analysis (PCA) to the spectra of each of the 18 Seyfert 1-like objects observed more than 15 times by the international ultraviolet explorer (IUE) from 1978 until the end of 1991.
PCA allows us to decompose the \Lya\ and \civl\ emission-lines of each object into two components with uncorrelated variations.
We find that the principal component describes correlated continuum and line variations, whereas the rest component reveals the line-parts that do not vary in tune with the continuum.
A cross-correlation analysis in NGC 5548 reveals that variable line features in the rest component follow the continuum by about 25 days.
The symmetry usually observed in the component's line profiles excludes that the velocity field of the line-emitting region is dominated by radial motion.
In some objects, the principal component reveals a clearly double-peaked line profile.

The results suggest that the variable line feature in the principal and the rest component is respectively emitted in the inner and outer parts of an extended broad-line region (BLR).
The fraction of the line that varies together with the continuum seems to decrease significantly with increasing luminosity.
We interpret this trend as a consequence of a larger BLR in more luminous objects.
Finally, we propose a solution to the \civ/\Lya\ ratio problem without referring to optically thin emission-line clouds.
\keywords{Line: profiles -- Galaxies: active -- quasars: emission lines -- Galaxies: Seyfert -- Ultraviolet: galaxies}
\end{abstract}

\section{\label{intro}Introduction}

Intensive studies of some Seyfert 1 galaxies have shown that emission-line variations are clearly a response to continuum variations (e.g. NGC 5548: Clavel et al. 1991; NGC 3783: Reichert et al. 1994; Fairall 9: Rodr\'{\i}guez-Pascual et al. 1997).
This gives strong support to the standard idea that emission-lines in active galactic nuclei (AGN) arise from photoionization of many small dense clouds moving around a central ionizing source (Netzer 1990).
The nature of these clouds is still unclear, but an attractive possibility is that they might be the envelopes of bloated stars (Alexander 1997 and references therein).

To understand the structure and the kinematics of the emission-line region, it is of great interest to be able to decompose the emission-lines into several independent components.
Most previous attempts decomposed the line profile into several Gaussian components (e.g. Wamsteker et al. 1990).
Such decompositions do not ensure that the components are really varying independently.

An interesting way to overcome this problem is to decompose the emission-lines by a principal component analysis (PCA).
This was first done by Mittaz et al. (1990), who applied a PCA to the IUE spectra of NGC 4151.
They concluded that PCA has only a fairly limited application to the analysis of spectral variability, because they could not find the physical meaning of most components.
The analysis we propose here is similar except that we consider only two components: the principal component and the rest component, which comprises all minor variations.
Doing so, the physical interpretation of the components becomes much easier and PCA turns out to be a promising way to study emission-line variations and hence the structure and the kinematics of the line emitting region.

The great advantage of the PCA, as compared to reverberation mapping (Peterson 1993) or cross-correlation methods, is that it does not require very well-sampled observations.
PCA allows us to study objects that have much too sparse observations to perform a meaningful cross-correlation between continuum and line fluxes.
Our sample of 18 objects is to our knowledge the greatest on which line profile variations were studied in a consistent way.
It gives us the opportunity to identify common behaviors among the AGN diversity and to look for possible trends with the object luminosity.

\section{\label{obs}Observational material}

All spectra were down-loaded from the uniform low dispersion archive (ULDA) version 4.0 in the Swiss ULDA national host in Lausanne.
This database contains almost all the low dispersion (resolution\,$\sim$\,6\,\AA) ultraviolet (UV) spectra obtained by the IUE satellite from 1978 to the end of 1991 (Courvoisier \& Paltani 1992).
The spectra of the ULDA database are processed with the IUESIPS reduction software described by Wamsteker et al. (1989) and references therein.
We are aware that the superior NEWSIPS reduction software would improve the signal-to-noise ratio of the spectra.
However, the IUE final archive (IUEFA) does not yet contain all the spectra used in this work and we considered that the improvement was not worth repeating the reduction for the more than thousand spectra used in this work.

We considered the 18 Seyfert 1-like objects (quasars and Seyfert galaxies) with more than 15 good quality spectra observed through the large aperture ($10''\times 20''$) of the short wavelength prime (SWP) camera.
Each spectrum was viewed separately and we excluded sky and noisy spectra, as well as some spectra with bright spots.
The redshift of 3C 273 is so that the \civl\ line coincides with a reseau mark at \Lobs\,1792--1796\,\AA, its most spoiled spectra were also excluded.

Because of the IUE wavelength scale uncertainty due to small displacements of the target within the IUE aperture, we realigned all spectra according to the peak of the geocoronal \Lyal\ emission fitted by a Gaussian.
This was done for all objects except NGC 4151, because the geocoronal emission is contaminated by its small redshifted (z\,=\,0.0033) \Lya\ line.
We choose the geocoronal emission rather than emission-lines from the object to avoid problems with double-peaked profiles and to use an identical procedure for all objects.
The dispersion of the shifts that we measured is $\sim$\,1.5\,\AA, but shifts up to $\sim$\,5\,\AA\ are observed.
This is enough to introduce spurious antisymmetric line profile variations (Türler \& Courvoisier 1997).
Finally, the spectra were rebinned into 1\,\AA\ bins for computational convenience.
Throughout this paper, the flux density is expressed in $10^{-14}\,\ergscmA$ and the integrated line flux is expressed in $10^{-14}\,\ergscm$.

\section{\label{pca}Principal component analysis}

PCA is a mathematical tool, which can reduce a multi-dimensional data set into a small number of linearly independent variables.
A general description of the PCA method can be found in Kendall \& Stuart (1976) or Jolliffe (1986).
The special case of PCA applied to AGN spectra is described by Mittaz et al. (1990) or by Francis et al. (1992) for object-to-object variations.

\begin{figure}[tb]
\begin{center}
\psfig{file=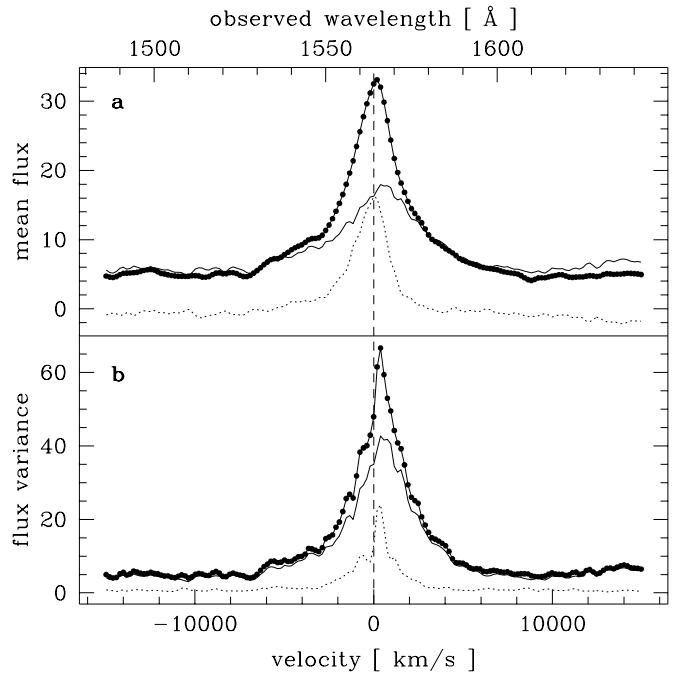,width=9cm}
\caption{\label{fig1}Principal component analysis of the \civl\ line in NGC 3783. \textbf{a} decomposition of the mean spectrum (points) into the principal component (solid line) and the rest component (dotted line). \textbf{b} similar decomposition of the variance spectrum. The flux is expressed in $10^{-14}\,\ergscmA$}
\end{center}
\end{figure}

\subsection{\label{method}Description of the method}

The analysis we applied to the spectra of each object was done as follows.
We constructed a $m \times n$ matrix $\mathcal{F}$ containing the fluxes of the $m$ observations in the $n$ 1\,\AA\ wavelength bins $\lambda_i$ ($i=1\ldots n$) between \Lobs\,1229 and 1948\,\AA.
These 720 bins represent nearly the entire SWP range except the geocoronal \Lya\ emission bellow 1229\,\AA.
The fluxes in each $\lambda_i$ form a base $\{f(\lambda_i)\}$ of a $n$-dimensional space, in which the $m$ observations are represented by $m$ points.

The aim of the PCA is to form a new base $\{f(\lambda_i^{\prime})\}$, in which only a few number of the $n$ vectors $f(\lambda_i^{\prime})$ are relevant to describe the variations.
$\{f(\lambda_i^{\prime})\}$ is the base in which the expression of the covariance matrix $\mathcal{C}$ of $\mathcal{F}$ is diagonal.
The vectors $f(\lambda_i^{\prime})$ are thus the eigenvectors of $\mathcal{C}$.
They are called components, since they describe linearly independent forms of variations and can be represented as spectra (each component is a linear combination of the $n$ initial $f(\lambda_i)$).
For a given component $k$, the eigenvalue of $\mathcal{C}$ is equal to the flux variance $\Sig_k^2$ in this component.
The contribution of the component $k$ to the total variability is given by its relative importance $I_k$ defined by: $I_k=\Sig_k^2 /\sum_{i=1}^{n}{\Sig_i^2}=\Sig_k^2/\mbox{Trace}(\mathcal{C})$.
The principal component is thus the component with the highest $I_k$ and its direction in the $n$-space follows the maximal elongation of the $m$ observation points.

The matrix $\mathcal{F}^{\prime}$, which is the expression of $\mathcal{F}$ in the new base $\{f(\lambda_i^{\prime})\}$, contains the fluxes of the $m$ observations in the $n$ ``pseudo-wavelengths'' $\lambda_i^{\prime}$.
We therefore have the $n$ lightcurves of the components from which we derive the mean flux $\mean{f}_i$ and the flux variance $\Sig_i^2$ in each component $i$.
The raw components are normalized to unity and their sign is undefined.
In order to be able to add or subtract components, we have to normalize them properly.
Two possible additive normalizations are (a) to multiply the components by their mean flux $\mean{f}_i$ or (b) to multiply the squared components by their flux variance $\Sig_i^2$.
The total component (i.e. the sum of all components) is then equal to (a) the mean spectrum or (b) the variance spectrum (i.e. respectively the mean flux and the flux variance in each $\lambda_i$).

In our analysis, we concentrate on only two components: the principal component and the rest component, which is constructed by subtracting the principal from the total component and is therefore the sum of the $n-1$ minor components.
In Fig. \ref{fig1}, we show the decomposition into these two components of (a) the mean and (b) the variance spectrum taken around the \civl\ line of NGC 3783.
The lightcurve of the total component is obtained by a vectorial addition of the lightcurves in the $n$ components (i.e. for each observation the squared flux in the total component is the sum of the squared fluxes in the $n$ components).

\begin{figure}[tb]
\begin{center}
\psfig{file=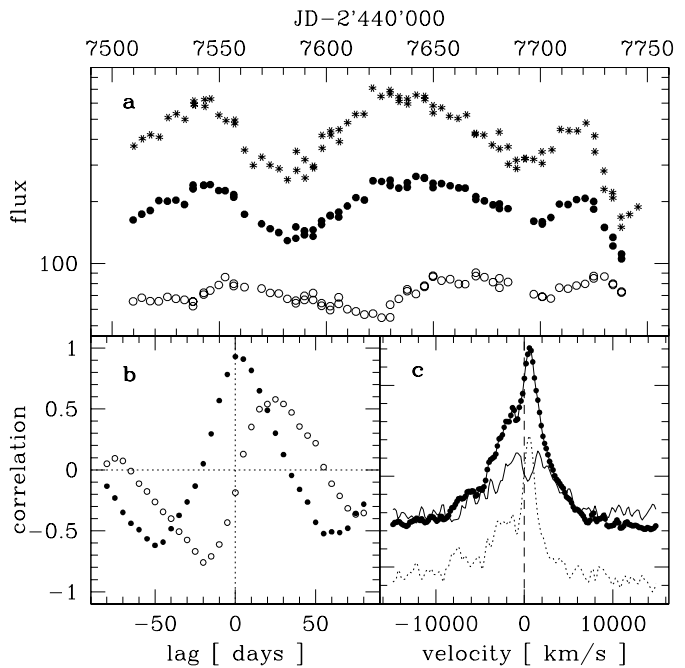,width=9cm}
\caption{\label{fig2}PCA study of the 1989 monitoring campaign on NGC 5548. \textbf{a} logarithmic lightcurves of the continuum flux at \Lrest\,1350\,\AA\ from Clavel et al. (1991) (star points), of the principal component (filled points) and of the rest component (open points). \textbf{b} cross-correlations of the principal component (filled points) and the rest component (open points) with the continuum lightcurve at \Lrest\,1350\,\AA. \textbf{c} PCA decomposition of the \civ\ line as in Fig. \ref{fig1}a}
\end{center}
\end{figure}

\subsection{\label{properties}Properties of the components}

The power of the PCA lies in the diagonalization of the covariance matrix $\mathcal{C}$ of $\mathcal{F}$.
A diagonal covariance matrix ensures that all structures constituting a single PCA component vary simultaneously, while the different components have variations that are completely uncorrelated at zero lag.

We applied the PCA on the widest possible spectral range to include a significant fraction of the continuum in the analysis.
This ensures that the principal component reflects the main continuum variations, since the continuum is known to vary with larger amplitudes than the emission-lines.
Figure \ref{fig1}b illustrates that in NGC 3783 nearly all continuum variations are described by the principal component alone.
Under this assumption, we can physically interpret the line profiles in the two components.
The line profile in the principal component shows which line-part varies simultaneously with the continuum, whereas the line profile in the rest component shows line-parts that do not vary in tune with the continuum. 

To test this interpretation and to further investigate the relationship between the continuum and our two components, we applied the PCA method to only the well sampled observations of the 1989 monitoring campaign on NGC 5548 (Clavel et al. 1991).
In Fig. \ref{fig2}a, we compare the principal and the rest component lightcurves with the original continuum lightcurve at \Lrest\,1350\,\AA\ from Clavel et al. (1991) in units of $10^{-16}\,\ergscmA$.

We analysed the correlation of the two components with the UV continuum using the interpolated cross-correlation function (ICCF) of Gaskell \& Sparke (1986) as described by White \& Peterson (1994).
The results show that the principal component variations are indeed strongly correlated with those of the continuum flux at zero lag, while the rest component variations are completely uncorrelated at zero lag with those of the continuum.
However, the cross-correlation between the rest component and the UV continuum suggests that the variable line-parts in the rest component follow the continuum with a delay of the order of 25 days.
This lag is twice as large as the lag between the continuum and the integrated \Lya\ and \civ\ lines (8--16 days) (Clavel et al. 1991), which means that the PCA divides the line into two parts: a line-part that responds to the continuum with a very small delay (\ltsim 5 days) and a line-part that responds with a much longer delay ($\sim$\,25 days).

The \civ\ line profile of these two parts is shown in Fig. \ref{fig2}c.
The full width at half maximum (FWHM) of the line-part in the rest component is much narrower (1\,850\,$\kms$) than in the principal component (7\,300\,$\kms$).
However, the full width at 1/4 maximum of the line-part in the rest component (6\,650\,$\kms$) shows that high velocities are also present.

This analysis suggests that the line-part in the principal component is emitted in the inner broad-line region (BLR), since it responds with large amplitudes and nearly zero lag to the continuum, whereas the line-part in the rest component is both constituted of a less varying contribution from outer parts of the BLR that respond with a greater delay to the continuum and of a non-varying contribution from the narrow-line region (NLR).

The lag of $\sim$\,25 days found for the rest component in NGC 5548 is comparable to the lag of the \ciiil\ line (26--32 days) (Clavel et al. 1991) and to the Balmer-line lags ($\sim$\,20 days for \Hb) (e.g. Peterson et al. 1991).
It is therefore possible that the \civ\ line-part in the rest component is emitted at about the same distance from the ionizing continuum than the \ciii\ or the \Hb\ line.

In other objects, the temporal sampling of the observations does not allow us to determine a clear lag between the two components and the continuum.
However, this will be possible with the other monitoring campaign data, that will soon be available in the IUEFA.

\section{\label{para}Definition of the variability parameters}

From the PCA results, it is possible to derive line and continuum variability parameters.
The mean continuum $\mean{f\lcont}_{k}$ and line $\mean{f\line}_{k}$ fluxes are derived from the component $k$ normalized by its mean flux $\mean{f}_{k}$ (Fig. \ref{fig1}a).
The continuum $\Sig\lcont$ and line $\Sig\line$ flux dispersions are estimated from the principal component normalized by its flux dispersion $\Sig\prin$.
This derivation is motivated by the theoretical argument exposed in Sect. \ref{theo}.
The determination of the continuum and the method of line integration is described in Sect. \ref{deter}.

\subsection{\label{theo}Theoretical basis}

In the following, the temporal average \mean{f(\lambda)} and the temporal dispersion $\Sig(\lambda)$ of the $n$ spectra $f(\lambda, t_i)$ measured at time $t_i$ ($i=1\ldots n$) are defined as
\begin{eqnarray}
\label{eq1}\mean{f(\lambda)}&=&\frac{1}{n}\sum_{i=1}^{n}f(\lambda,t_i)\qquad\qquad\mbox{and}\\
\label{eq2}\Sig^2(\lambda)&=&\frac{1}{n-1}\sum_{i=1}^{n}(\,\Delta f(\lambda,t_i)\,)^2\,,
\end{eqnarray}
where $\Delta f(\lambda,t_i)\equiv f(\lambda,t_i)-\mean{f(\lambda)}$.
At each $t_i$, the total spectrum is the sum of a continuum and a line contribution $f\tot(\lambda,t_i)=f\cont(\lambda,t_i)\,+\,f\line(\lambda,t_i)$.
This relation combined with Eqs. (\ref{eq1}) \& (\ref{eq2}) gives
\begin{eqnarray}
\label{eq3}\protect\mean{f\tot(\lambda)}&=&\protect\mean{f\cont(\lambda)}\,+\,\protect\mean{f\line(\lambda)}\\
\Sig\tot^2(\lambda)&=&\Sig\cont^2(\lambda)\,+\,\Sig\line^2(\lambda)\nonumber\\
\label{eq4}& & +\,\frac{2}{n-1}\sum_{i=1}^{n}\Delta f\cont(\lambda,t_i)\cdot \Delta f\line(\lambda,t_i)
\end{eqnarray}
Eq. (\ref{eq4}) shows that contrary to the mean spectrum, the variance spectrum can not be simply decomposed into a continuum and a line contribution.

It would however be possible to decompose the dispersion spectrum (i.e. the flux dispersion in each $\lambda_i$) into a continuum and a line contribution if
\begin{equation}\label{eq5}
\Delta f\line(\lambda,t_i)=R(\lambda)\cdot\Delta f\cont(\lambda,t_i)\,,
\end{equation}
because this would imply that $\Sig\line(\lambda)=R(\lambda)\cdot\Sig\cont(\lambda)$ via Eq. (\ref{eq2}) and thus that Eq. (\ref{eq4}) reduces to
\begin{eqnarray}
\Sig\tot(\lambda)&=&\sqrt{\Sig\cont^2(\lambda)\,+\,\Sig\line^2(\lambda)\,+\,2R(\lambda)\Sig\cont^2(\lambda)}\nonumber\\
\label{eq6}&=&\Sig\cont(\lambda)\,+\,\Sig\line(\lambda)\,.
\end{eqnarray}

Equation (\ref{eq5}) would correspond to substitute a time independent response function $R(\lambda)$ for actual convolution with the transfer function.
Since there is a significant lag between the line and the continuum variations, Eq. (\ref{eq5}) is incorrect and thus we can not derive the line variability $\Sig\line$ from the dispersion spectrum via Eq. (\ref{eq6}).

However, we have shown in Sect. \ref{properties} that the principal component is correlated with the continuum at zero lag.
This is not only verified in NGC 5548, but is a direct consequence of the way we applied the PCA.
Therefore, Eq. (\ref{eq5}) is correct for the line-part in the principal component and thus, according to Eq. (\ref{eq6}), we can estimate the line variability $\Sig\line$ by integrating the line-part in the principal component normalized by its flux dispersion $\Sig\prin$.
This estimation gives a lower limit to the actual line variability, since it does not consider the delayed line variations, but has the advantage of excluding spurious variability due to noisy spectra.

\begin{table*}[tb]
\caption{\label{tab1}List of the objects in our sample with some characteristic parameters derived from the PCA. log(L) is the logarithm of the luminosity expressed in $\ergsA$, the FWHM is expressed in \kms. Other quantities have no units. The parameters are defined in Sect. \ref{global}}
\begin{flushleft}
\begin{tabular}{lrlcccrrrrrr@{}}
\hline
\rule[-1.0em]{0pt}{2.5em}Object& \multicolumn{1}{c}{n}& \multicolumn{1}{c}{z}& \multicolumn{1}{c}{log(L)}& I$\prin$& $\frac{\mbox{$\Sig\rest$}}{\mbox{$\Sig\prin$}}$& \multicolumn{2}{c}{$\Et$}& \multicolumn{2}{c}{FWHM$\prin$}& \multicolumn{2}{c@{}}{FWHM$\rest$}\\
\rule[-0.1em]{0pt}{1.1em}& & & & & & \multicolumn{1}{c}{\Lya}& \multicolumn{1}{c}{\civ}& \multicolumn{1}{c}{\Lya}& \multicolumn{1}{c}{\civ}& \multicolumn{1}{c}{\Lya}& \multicolumn{1}{c@{}}{\civ}\\
\hline\rule{0pt}{1.2em}Ark 120& 33& 0.0330& 41.39& 0.70& 0.20& 0.61$\pm$0.06& 0.37$\pm$0.07& 5\,700& 8\,650& 3\,800& 3\,450\\
3C 120& 21& 0.0336& 40.68& 0.48& 0.24& 0.42$\pm$0.07& 0.42$\pm$0.04& 1\,950& 2\,300& 3\,800& 2\,250\\
3C 273& 110& 0.158& 43.26& 0.85& 0.12& 0.21$\pm$0.03& 0.17$\pm$0.10& 14\,450& ---& 3\,150& 3\,900\\
3C 382& 22& 0.0578& 41.34& 0.88& 0.11& 0.56$\pm$0.05& 0.57$\pm$0.03& 8\,650& 13\,100& 7\,500& 10\,550\\
3C 390.3& 23& 0.0561& 40.74& 0.50& 0.31& 0.47$\pm$0.05& 0.87$\pm$0.06& 8\,550& 8\,200& 2\,500& 1\,400\\
ESO 141-55& 26& 0.0368& 41.67& 0.85& 0.12& 0.88$\pm$0.03& 0.55$\pm$0.05& 4\,950& 5\,750& 3\,500& 5\,250\\
Fairall 9& 84& 0.046& 41.77& 0.86& 0.21& 0.85$\pm$0.02& 0.62$\pm$0.03& 4\,200& 3\,950& 2\,150& 2\,050\\
GQ Comae& 17& 0.1653& 41.93& 0.66& 0.18& 0.40$\pm$0.05& 0.38$\pm$0.11& 4\,750& 5\,000& 3\,100& 2\,650\\
Mrk 279& 23& 0.0297& 41.22& 0.89& 0.11& 0.78$\pm$0.02& 0.53$\pm$0.03& 7\,700& 4\,200& 3\,700& 7\,000\\
Mrk 335& 28& 0.025& 41.17& 0.48& 0.50& 0.82$\pm$0.03& 0.78$\pm$0.08& 2\,850& 3\,700& 1\,600& 2\,900\\
Mrk 509& 37& 0.0355& 41.60& 0.70& 0.16& 0.78$\pm$0.03& 0.53$\pm$0.03& 6\,650& 5\,350& 3\,300& 2\,400\\
Mrk 926& 18& 0.048& 41.54& 0.85& 0.14& 0.43$\pm$0.03& 0.61$\pm$0.02& 7\,600& 10\,650& 5\,000& 5\,050\\
NGC 3516& 30& 0.009& 39.73& 0.81& 0.18& 0.71$\pm$0.02& 0.60$\pm$0.01& 6\,100& 8\,900& 2\,700& 2\,300\\
NGC 3783& 55& 0.0096& 40.24& 0.87& 0.18& 0.97$\pm$0.03& 0.60$\pm$0.01& 7\,400& 4\,650& 2\,300& 2\,200\\
NGC 4151& 314& 0.0033& 39.60& 0.89& 0.13& 0.94$\pm$0.08& 0.82$\pm$0.01& 8\,000& 7\,650& ---& 1\,850\\
NGC 4593& 25& 0.0087& 39.58& 0.76& 0.17& 0.79$\pm$0.04& 0.68$\pm$0.03& 10\,300& 9\,250& 3\,800& 3\,500\\
NGC 5548& 154& 0.017& 40.71& 0.78& 0.17& 0.91$\pm$0.01& 0.63$\pm$0.01& 6\,900& 7\,450& 1\,500& 1\,850\\
NGC 7469& 20& 0.0165& 40.77& 0.51& 0.42& 0.27$\pm$0.04& 0.33$\pm$0.05& 7\,000& 7\,500& 2\,000& 2\,850\\
\hline
\end{tabular}
\end{flushleft}
\end{table*}

\subsection{\label{deter}Continuum and emission-line determination}

On the basis of Eqs. (\ref{eq3}) \& (\ref{eq6}), we can derive both the mean and the dispersion of the line flux by integrating the emission-line above a defined continuum in the appropriate component.
To do this, we first defined five 30\,\AA\ bands that are usually free of emission or absorption features in the rest frame of the source at \Lrest\,1120--1150\,\AA, 1320--1350\,\AA, 1430--1460\,\AA, 1700--1730\,\AA, 1810--1840\,\AA.
The two last bands are both outside of the component domain \Lobs\,1229--1948\,\AA\ in 3C 273 and GQ Comae, due to their higher redshift.
For these two objects, we had to define a new band from \Lrest\,1665\,\AA\ to the end of the component, despite the possible \heiil\ and \oiiil\ contamination at those wavelengths.
Doing so, there are always four bands that constrain the continuum, whatever the redshift of the object is.

The continuum is estimated by fitting a straight line through the points in these four bands.
We used the ordinary least-squares regression ``OLS(X$\mid$Y)'' of Isobe et al. (1990) with its uncertainties on the slope and on the intercept.
To take into account systematic errors, the uncertainty on the continuum is assumed to be twice the uncertainty on the fit that was calculated, as all uncertainties in this paper, according to the general formula
\begin{equation}\label{eq7}
\Sig_f=\sqrt{\sum_{i=1}^{n}\left(\frac{\partial f}{\partial x_i}\Sig_{x_i}\right)^2}\,,
\end{equation}
where $f$ is a function of the $n$ variables $x_i$ with uncertainties $\Sig_{x_i}$.

The \Lya\ and \civ\ emission-lines are integrated above the continuum in the velocity ranges [\,$-$15\,000\,;\,$+$\,15\,000\,] and [\,$-$15\,000\,;\,$+$\,12\,000\,] respectively, as far as enabled by the redshift of the object.
We choose these ranges in order to include the \nvl\ line in the \Lyal\ integration and to exclude the \heiil\ and \oiiil\ lines from the \civl\ integration.
In the following, the \Lya\ line actually refers to the \Lya\,$+$\,\nv\ line.
The uncertainty on the line contribution is determined by integrating the line above the fitted continuum plus and minus its uncertainty.
The continuum parameters are arbitrary defined at \Lrest\,1350\,\AA.

\begin{figure}[tb]
\begin{center}
\psfig{file=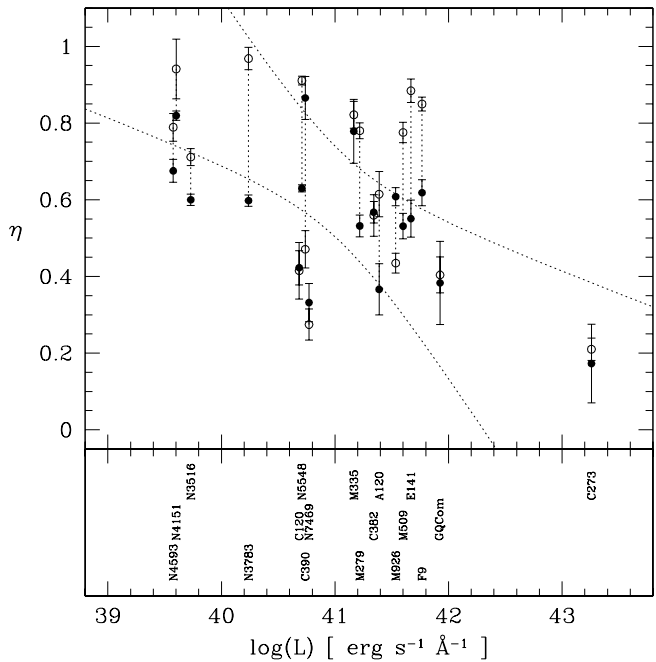,width=9cm}
\caption{\label{fig3}Fraction of the emission-line present in the principal component as a function of the object luminosity.
Open and filled points represent respectively \Lya\ and \civ\ lines. 
The two dotted curves represent the 3-$\sigma$ uncertainty on the bisector linear regression of Isobe et al. (1990)}
\end{center}
\end{figure}

\begin{figure}[tb]
\begin{center}
\psfig{file=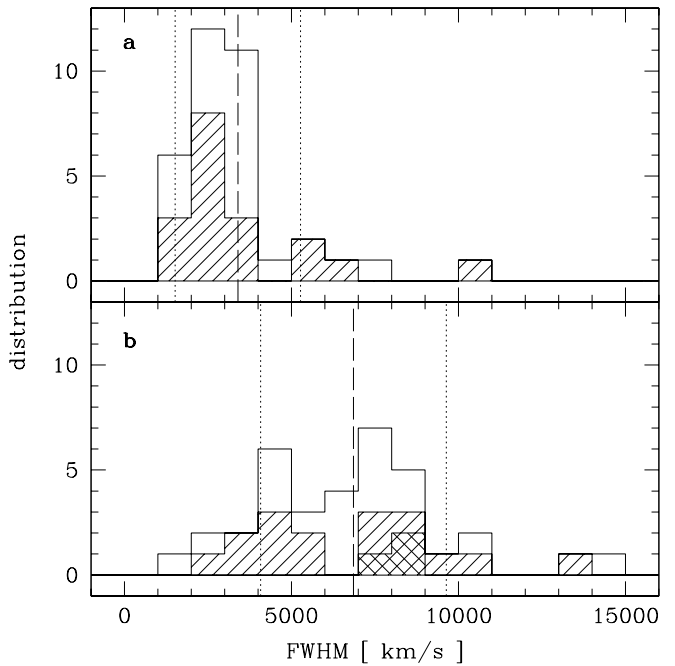,width=9cm}
\caption{\label{fig4}Distribution of the FWHM of the emission-lines in the rest component (\textbf{a}) and the principal component (\textbf{b}).
The \civ\ lines are shaded or twice shaded if they are double-peaked.
The mean (dashed line) and the dispersion (dotted lines) of the distribution are shown}
\end{center}
\end{figure}

\section{\label{resu}Results}

In the appendix, we show the PCA decomposition of the \civ\ and \Lya\ lines for all objects.
If the component's line profiles are similar for the \civ\ and the \Lya\ line of a same object, they can be very different from one AGN to the other.
This great diversity forced us to consider only very simple parameters to quantify the line properties of the components in order to find significant trends.
The general results and possible trends with luminosity are presented in Sect. \ref{global}, whereas comments on individual objects are given in Sect. \ref{individual}.

\subsection{\label{global}Global trends}

Tables \ref{tab1} \& \ref{tab2} display the most meaningful quantities derived from our analysis together with the number $n$ of considered spectra and the redshift $z$ of the object as given by the set of identifications, measurements and bibliography for astronomical data (SIMBAD).
log(L) is the logarithm of the luminosity density at \Lrest\,1350\,\AA\ expressed in $\ergsA$.
It was derived from the fitted continuum assuming a Hubble constant of $H_{0}=50\,\Hubble$.

To quantify the variability in the principal component with respect to other variations, we choose two parameters: $I\prin$ and $\Sig\rest/\Sig\prin$, which are respectively the relative importance of the principal component and the ratio of the flux dispersions in the rest and in the principal component, as defined in Sect. \ref{method}.
Both parameters show that in general the dominant variations are well described by the principal component alone ($I\prin$\gtsim 50\,\% and $\Sig\rest/\Sig\prin$\ltsim 50\,\%).

According to the discussion in Sect. \ref{properties}, the fraction $\Et$ of the line that varies together with the continuum is given by the ratio of the mean line fluxes in the principal and in the total component $\Et=\mean{f\line}\prin/\mean{f\line}\tot$.
Its dependence with luminosity shown in Fig. \ref{fig3} has a slope of $-0.28\pm0.06$/decade.
Here, as in the following, the slope was determined by using the ``OLS bisector'' unweighted linear regression of Isobe et al. (1990), as advised by these authors and the trend is highlighted in the figure by the 3-$\sigma$ uncertainty on the regression.
We applied a Spearman's test (e.g. Bevington 1969) to determine whether the correlation suggested by Fig. \ref{fig3} is significant or not.
The probability that such a correlation could occur by chance is 1.6\,\%.

We quantified the width of the \Lya\ and the \civ\ lines by evaluating their FWHM both in the principal and in the rest component.
The FWHM values displayed in Table \ref{tab1} are expressed in \kms\ with an uncertainty on the measure of $\pm$100\,\kms.
Some noisy components had to be smoothed to determine properly their width and in low redshift objects, for which the blue side of the \Lya\ line is not in the component, the FWHM was extrapolated from the half width at half maximum.
Even so, we could not determine a meaningful width for the \Lya\ line in the rest component of NGC 4151, because of geocoronal \Lya\ contamination and for the \civ\ line in the principal component of 3C 273, because there is nearly no line feature in this component.

The distribution of the FWHM in Fig. \ref{fig4} shows that there is no obvious difference between the \civ\ and the \Lya\ line.
The average width of the two lines is twice as large in the principal component (6\,850\,\kms) as in the rest component (3\,400\,\kms).
Individually, the line is always broader in the principal component than in the rest component, except for the \Lya\ line in 3C 120 and the \civ\ line in Mrk 279.
This clear result shows that the line-part that varies with the continuum is broader in general than the line-part that varies less.
In many objects however, the line in the rest component is broader than the typical width (300--1\,000\,\kms) of a line emitted in the NLR (Netzer 1990).

The \civ\ line profile in the principal component is clearly double-peaked in 3C 390.3, NGC 3516 and NGC 4151.
A double-peaked profile is known to be the signature of a thin rotating disk viewed close to edge-on (Welsh \& Horne 1991).
Recently, Goad \& Wanders (1996) showed that double-peaked profiles can also originate due to a non-uniform lighting of a spherical BLR by a predominantly biconical continuum emission.
Their model is able to reproduce a wide range of observed profiles and predicts that the FWHM should be generally larger in double- than in single-peaked profiles.
The FWHM that we measured for our three \civ\ double-peaked profiles ($\sim$\,8\,000\,\kms) are among the highest in the sample, in good agreement with their prediction.
The line profile in Mrk 509 can be seen as the transition between double- and single-peaked profiles.

\begin{table}[tb]
\caption{\label{tab2}The relative variability of the continuum measured at \Lrest\,1350\,\AA\ and of the integrated \Lya\ and \civ\ lines}
\begin{flushleft}
\begin{tabular}{lccc@{}}
\hline
\rule[-0.7em]{0pt}{2.0em}Object& $(\Sig/\mean{f})_{1350}$& $(\Sig/\mean{f})_{\mbox{\Lya}}$& $(\Sig/\mean{f})_{\mbox{\civ}}$\\
\hline\rule{0pt}{1.2em}Ark 120& 0.22$\pm$0.01& 0.11$\pm$0.01& 0.06$\pm$0.01\\
3C 120& 0.27$\pm$0.02& 0.09$\pm$0.02& 0.09$\pm$0.01\\
3C 273& 0.23$\pm$0.00& 0.04$\pm$0.01& 0.03$\pm$0.02\\
3C 382& 0.54$\pm$0.02& 0.24$\pm$0.02& 0.25$\pm$0.01\\
3C 390.3& 0.52$\pm$0.04& 0.15$\pm$0.02& 0.27$\pm$0.02\\
ESO 141-55& 0.27$\pm$0.01& 0.21$\pm$0.01& 0.13$\pm$0.01\\
Fairall 9& 0.58$\pm$0.01& 0.40$\pm$0.01& 0.29$\pm$0.02\\
GQ Comae& 0.53$\pm$0.03& 0.16$\pm$0.02& 0.15$\pm$0.04\\
Mrk 279& 0.41$\pm$0.01& 0.26$\pm$0.01& 0.18$\pm$0.01\\
Mrk 335& 0.16$\pm$0.01& 0.11$\pm$0.00& 0.11$\pm$0.01\\
Mrk 509& 0.21$\pm$0.01& 0.12$\pm$0.00& 0.08$\pm$0.01\\
Mrk 926& 0.57$\pm$0.02& 0.16$\pm$0.01& 0.23$\pm$0.01\\
NGC 3516& 0.48$\pm$0.01& 0.27$\pm$0.01& 0.23$\pm$0.01\\
NGC 3783& 0.46$\pm$0.01& 0.35$\pm$0.01& 0.22$\pm$0.01\\
NGC 4151& 0.69$\pm$0.02& 0.55$\pm$0.05& 0.48$\pm$0.01\\
NGC 4593& 0.43$\pm$0.02& 0.28$\pm$0.01& 0.24$\pm$0.01\\
NGC 5548& 0.35$\pm$0.01& 0.24$\pm$0.00& 0.17$\pm$0.00\\
NGC 7469& 0.18$\pm$0.01& 0.05$\pm$0.01& 0.05$\pm$0.01\\
\hline
\end{tabular}
\end{flushleft}
\end{table}

\begin{figure}[tb]
\begin{center}
\psfig{file=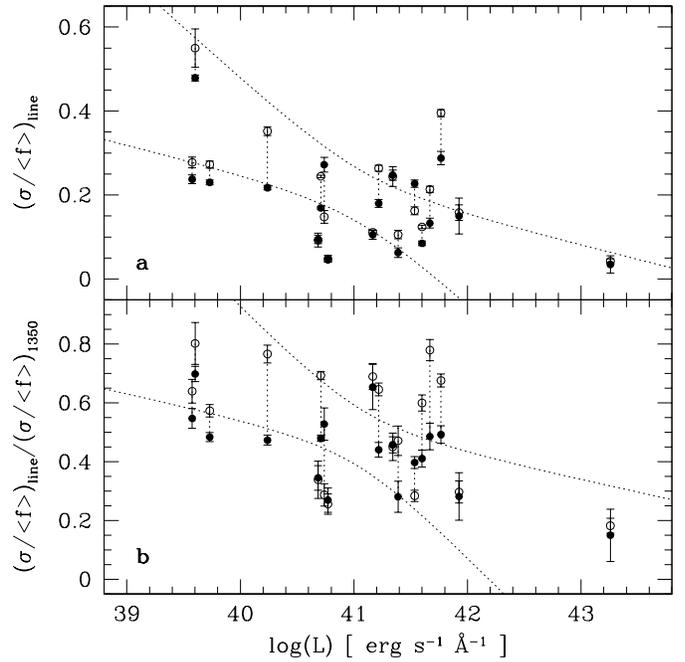,width=9cm}
\caption{\label{fig5}Luminosity dependence of the line variability (\textbf{a}) and a parameter describing the line-to-continuum response (\textbf{b}).
Open and filled points represent respectively \Lya\ and \civ\ lines. 
The dotted curves are defined as in Fig. \ref{fig3}}
\end{center}
\end{figure}

We quantified the variability of the continuum and of the \Lya\ and \civ\ lines by the ratio of the flux dispersion on the mean flux $\Sig/\mean{f}$.
This ratio corresponds to the parameter $F\var$ first defined by Clavel et al. (1991) to describe the flux variability.
We calculated the values shown in Table \ref{tab2} as explained in Sect. \ref{para}.
The luminosity dependence of the line variability shown in Fig. \ref{fig5}a has a slope of $-0.16\pm0.03$/decade.
The Spearman's test probability that such an anticorrelation could occur by chance is 1.8\,\%, but rises to $\sim$\,10\,\%, when we consider only the \Lya\ or the \civ\ line and reaches 70\,\% for the continuum variability.
This absence of correlation between the continuum variability and the luminosity is in contradiction with Paltani \& Courvoisier (1994), who found a strongly significant anticorrelation in a sample of 72 Seyfert-like objects with a slope of $-0.046$/decade.
The fact that this trend remains hidden to us with a sub-sample of 18 objects is most probably due to its weak slope and to the high scatter of the points.
This increases our confidence in the significant trends that we found for the two emission-lines.

The ratio of the line variability on the continuum variability $(\Sig/\mean{f})\line/(\Sig/\mean{f})_{1350}$ is perhaps even more meaningful, since it describes the line-to-continuum response.
Its luminosity dependence shown in Fig. \ref{fig5}b has a steeper slope ($-0.24\pm0.05$/decade) than that of the line variability, but the Spearman's test probability is similar (1.7\,\%) and the scatter of the points is higher.
The great similarity with the relation in Fig. \ref{fig3} is due to the fact that $\Et$ and $(\Sig/\mean{f})\line/(\Sig/\mean{f})_{1350}$ are closely related, as shown by the following theoretical argument.

Since the PCA results suggest that a fraction $\Et$ of the line varies in tune with the continuum, while the remaining fraction ($1-\Et$) is nearly constant, we can write
\begin{eqnarray}
\label{eq8}f\line(t_i)&\simeq&R\,[\,\Et\,f\lcont(t_i)+(1-\Et)\mean{f\lcont}\,]\\
\label{eq9}&\simeq&R\,[\,\mean{f\lcont}+\Et\,\Delta f\lcont(t_i)\,]\,,
\end{eqnarray}
where $R$ is a response constant and $\Delta f(t_i)\equiv f(t_i)-\mean{f}$.
We can now equal the first and the second term in Eq. (\ref{eq9}) with $\mean{f\line}$ and $\Delta f \line(t_i)$ respectively:
\begin{eqnarray}
\label{eq10}\mean{f\line}&\simeq&R\,\mean{f\lcont}\\
\label{eq11}\Delta f\line(t_i)&\simeq&R\,\Et\,\Delta f\lcont(t_i))\,.
\end{eqnarray}
In accordance with Eq. (\ref{eq2}), Eq. (\ref{eq11}) implies that $\Sig\line\simeq R\,\Et\,\Sig\lcont$, which together with Eq. (\ref{eq10}) leads to
\begin{equation}\label{eq12}
\Et\simeq\frac{(\Sig/\mean{f})\line}{(\Sig/\mean{f})\lcont}\,.
\end{equation}
This straightforward calculation shows that both $\Et$ and $(\Sig/\mean{f})\line/(\Sig/\mean{f})_{1350}$ describe in a way the line-to-continuum response.

In all but two objects (3C 390.3 and Mrk 926), the line response is stronger within the uncertainties for the \Lya\ than for the \civ\ line.
This naturally leads to a decrease of the \civ/\Lya\ ratio with increasing continuum flux, as it was observed in some Seyfert galaxies (Peterson 1993).
3C 390.3 was until now the only object in which the \civ/\Lya\ ratio was observed to increase with increasing continuum flux (Wamsteker et al. 1997).
Our results confirm this peculiar behavior in 3C 390.3 and predict a similar one in Mrk 926.

We did not quantify line profile asymmetries, but a look at the components displayed in the appendix shows that most of their line profiles are roughly symmetric.
The observed absence of strong line asymmetries in the components of most objects excludes that their line-emitting region is dominated by infall or outflow (Rosenblatt et al. 1994).
However, minor line asymmetries might reflect some radial motion within a predominantly chaotic or rotational velocity field.

\subsection{\label{individual}Comments on individual objects}

Apart from 3C 390.3, which has a different \civ/\Lya\ ratio behavior than most AGN (Sect. \ref{global}), other objects have some peculiarities.
The most luminous radio-loud quasar in our sample is 3C 273, its principal component has only a very broad and nearly undetectable line feature.
This leads to a very weak line variability, as was already noticed by Ulrich et al. (1993) for the \Lya\ line.
Since 3C 273 is the only well enough observed luminous object, it is difficult to know whether or not its blazar characteristics are responsible for this peculiarity.
However, the fact that 3C 273 is well inside the obtained luminosity trend suggests that other luminous quasars would also have nearly constant emission-lines.

Fairall 9, on the other hand, is an object which does not follow the general luminosity trend.
Its line variability behavior is typical for a Seyfert galaxy, whereas its luminosity is that of a quasar.
This was already pointed out by Rodr\'{\i}guez-Pascual et al. (1997), who found that the emission-line lags in Fairall 9 (\Lya: 14--20 days; \heiil: \ltsim\,4 days) are comparable to those in NGC 5548, despite the difference in luminosity of a factor of ten.
It suggests that the luminosity of Fairall 9 during the first years of IUE observations was extraordinary high and hence that its average luminosity is overestimated.

3C 120 and NGC 7469 are two other objects that are always outside of the 3-$\sigma$ uncertainty curves.
Their line variability is very small, because they are the two only objects with a clear asymmetric line profile in their first component.
The blue wing of the line is not correlated with the continuum in 3C 120, whereas it seems even anticorrelated in NGC 7469.
Such asymmetries are qualitatively consistent with an infalling BLR, since the blue wing is then emitted behind the continuum source having thus a higher emission-line lag than the red wing emitted in front of the source.
However, both objects are not well enough observed to draw strong conclusions only based on these PCA results.

\section{\label{discu}Discussion}

Perhaps the most important result of the PCA is that a great fraction of the emission-line does not vary together with the continuum.
This fraction has a line profile that is usually too broad to originate only in the NLR and thus the classical division of the emission-line into a variable BLR and a constant NLR contribution has to be reviewed.
The PCA decomposition gives support to the model proposed by Brotherton et al. (1994) in which the traditional BLR is divided in two components: a very broad line region (VBLR) and an intermediate line region (ILR), which have typical velocities of $\sim$\,7\,000\,\kms\ and $\sim$\,2\,000\,\kms, respectively.
The line profile in the principal component would then be the signature of the VBLR, whereas the rest component's profile would be the signature of the ILR plus a constant contribution from the NLR.

Under this assumption, the delay of $\sim$\,25 days that we measured between the rest component and the UV continuum in NGC 5548 (Sect. \ref{properties}) can be understood as the distance of the ILR.
This result is in good agreement with Brotherton et al. (1994), who estimate that the ILR is about 10 times more distant from the ionizing continuum than the VBLR.
We note however that the line profile in the rest component has wings which extend to significant fractions of those in the principal component.
This implies the presence of high velocity gas in the ILR and hence suggests that the VBLR and the ILR are not completely separated.

The ratio of the line on the continuum variability naturally decreases when the light-travel time across the line-emitting region $t\lt$ is of the order of the typical variability timescale of the ionizing continuum $t\var$.
If $t\lt\ll t\var$, the entire line-emitting region will respond as a whole to continuum variations, whereas if $t\lt\gg t\var$, different sub-regions will respond with different delays to the ionizing continuum averaging out the variations (Rosenblatt et al. 1994).
In our interpretation, the lines emitted in the ILR have weak variations, because the extended ILR implies that $t\lt\!>\!t\var$, whereas the lines emitted in the VBLR vary in tune with the continuum, because this region is small enough to satisfy $t\lt\!<\!t\var$.
A simple way to formally link the fraction $\Et$ of the line in the principal component (i.e. emitted in the VBLR) to the two characteristic timescales is to write
\begin{equation}\label{eq13}
\Et\simeq \exp{(-t\lt/t\var)}\,,
\end{equation}
where $t\lt$ is now the light-crossing time of the BLR.
According to Eq. (\ref{eq13}), we obtain with our data the following relation with luminosity
\begin{equation}\label{eq14}
t\lt/t\var\simeq -\ln{(\Et)}\propto L^{0.52\pm0.09}\,,
\end{equation}
which shows that the BLR size increases with respect to $c\,t\var$ as the luminosity increases.
If we assume that $t\lt$ is proportional to the observed delays between the line and the continuum and thus that it varies roughly with the luminosity as $t\lt\propto L^{0.5}$ (Kaspi et al. 1996), the relation in Eq. (\ref{eq14}) suggests that the variability timescale of the ionizing continuum $t\var$ is independent of the object luminosity.

The usually observed decrease of the \civ/\Lya\ ratio with increasing continuum flux was interpreted as being due to a population of optically thin clouds (Shields et al. 1995).
Such a model predicts that the broad emission-line component emitted in the inner BLR by optically thin clouds should be less variable than the narrower component emitted in the outer BLR by optically thick clouds.
This prediction seems to be verified for the \Hb\ line in some Seyfert galaxies (Mrk 590: Peterson et al. 1993; Mrk 335: Kassebaum et al. 1997), but was never observed to our knowledge for the \Lya\ or the \civ\ line.
Our result that the most varying line-part is usually also the broadest is in contradiction with such a model.

We propose that the \civ/\Lya\ ratio problem is due to a significant decrease of the variability timescale $t\var$ between the ionizing thresholds of H\,{\sc i} and \civ, respectively at 912\,\AA\ (13.6\,eV) and at 192\,\AA\ (64.5\,eV).
If we assume that the light-crossing time of the BLR $t\lt$ is similar for the two lines, this would lead to a better line-to-continuum response for the \Lya\ than for the \civ\ line (Eq. (\ref{eq12}) \& (\ref{eq13})), implying a decrease of the \civ/\Lya\ ratio with increasing continuum flux.

The lack of far UV observations does not allow us to test this interpretation.
However, there are evidences that the continuum varies more rapidly at higher energies than in the UV.
Nandra et al. (1991) found that the X-ray flux in NGC 5548 varies significantly on timescales of hours and that it can vary by a factor of two in less than a few days.
More recently, Marshall et al. (1997) found that the extreme UV (150--200\,eV) can vary by a factor of two on timescales of 0.5 day.
Such a decrease of the variability timescale toward shorter wavelength from the UV to the extreme UV domain is qualitatively in agreement with our interpretation.

\acknowledgements{We thank S. Paltani for precious help and advice on the use of PCA and we are very grateful to the anonymous referee for his detailed comments, which helped us strongly to improve this work.}

\appendix
\onecolumn
\section*{Appendix A: PCA decomposition of the \civ\ line}
PCA decomposition of the mean \civl\ line profile (points) into the principal component (solid line) and the rest component (dotted line).
The abbreviated name of the object is shown in each panel.
The zero velocity (dashed line) is defined by the redshift.
On the flux axis, the first big tick is always at zero flux and small ticks are at every $10^{-13}\,\ergscmA$.

\begin{figure}[htb]
\begin{center}
\psfig{file=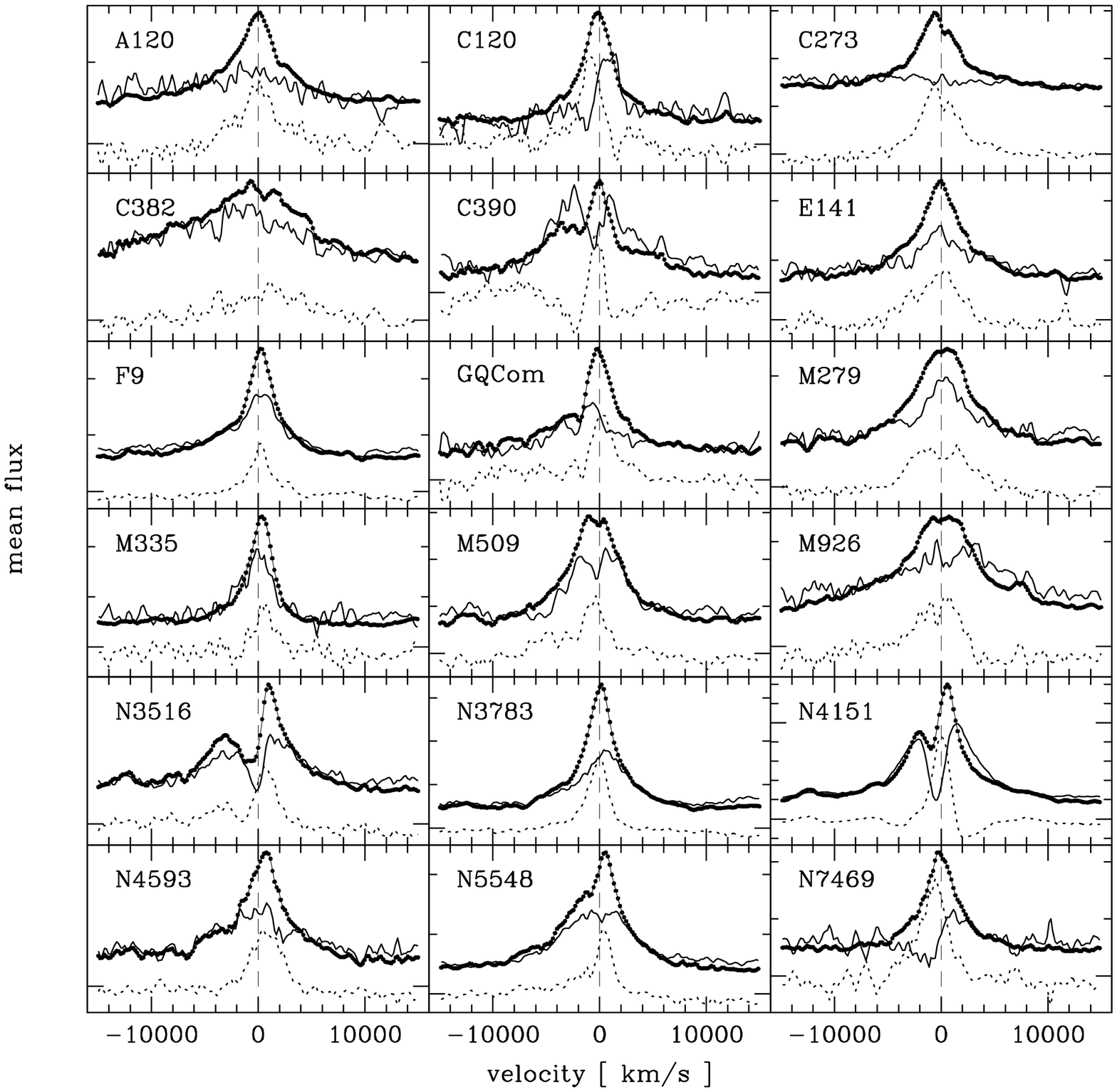,width=\textwidth}
\end{center}
\end{figure}

\section*{Appendix B: PCA decomposition of the \Lya\ line}

Same as in appendix A, but for the \Lya\ line.
The blue wing of the line could not always be analysed because of geocoronal \Lya\ contamination.

\begin{figure}[htb]
\begin{center}
\psfig{file=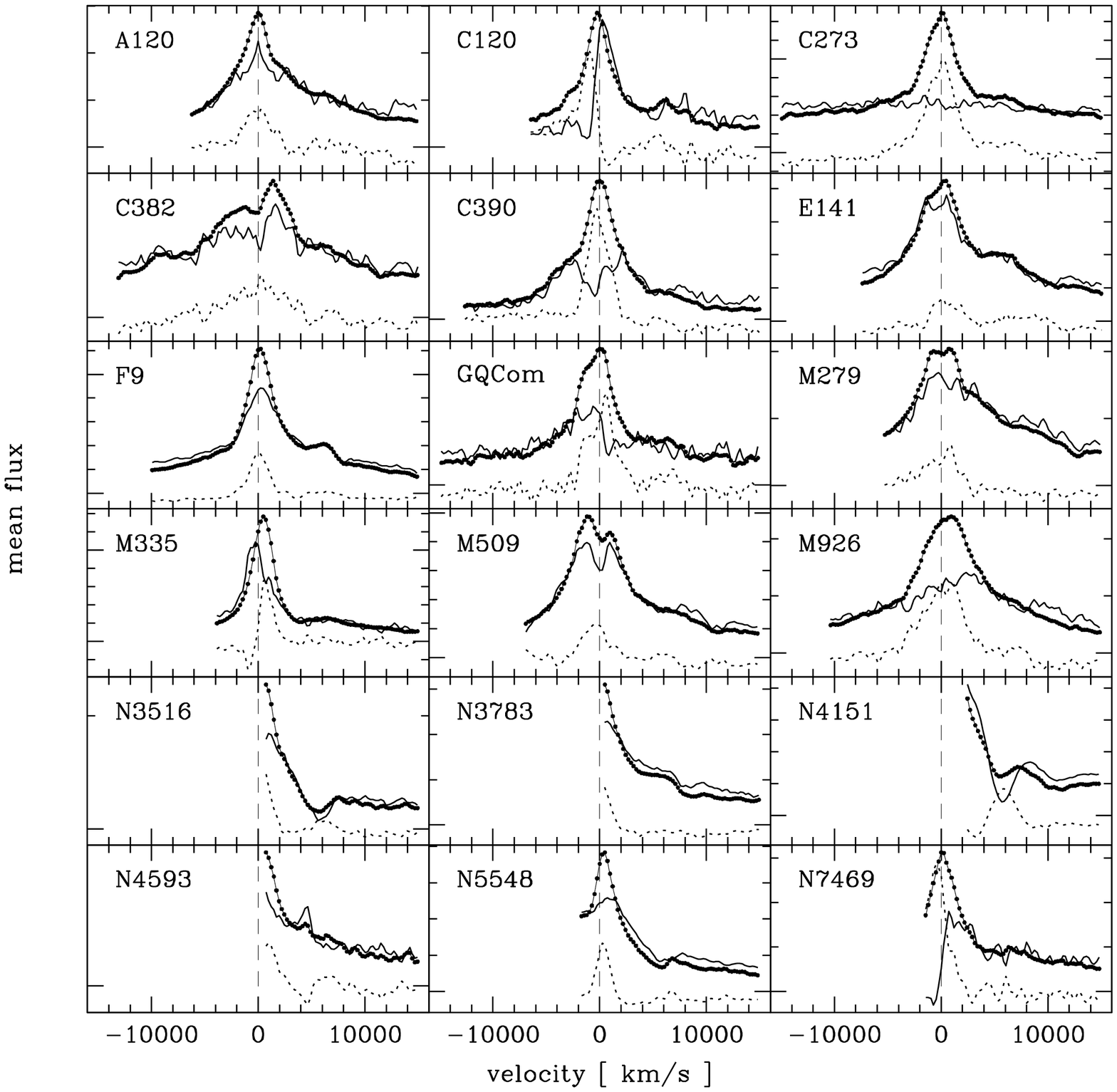,width=\textwidth}
\end{center}
\end{figure}

\end{document}